\documentclass{sig-alternate-hotpets}

\usepackage[utf8]{inputenc}
\usepackage{graphicx}
\usepackage{hyperref}

\newcommand{\bpf}{BPF\ }

\begin{document}
\title{Flexible Anonymous Network}
%\subtitle{[Modified for HotPETS]}

\numberofauthors{3} %  in this sample file, there are a *total*
\author{
% 1st. author
\alignauthor
Florentin Rochet\\
       \affaddr{UCLouvain Crypto Group}\\
 %      \email{florentin.rochet@uclouvain.be}
% 2nd. author
\alignauthor
Olivier Bonaventure\\
       \affaddr{UCLouvain IP Networking Lab}\\
 %      \email{olivier.bonaventure@uclouvain.be}
% 3rd. author
\alignauthor Olivier Pereira\\
       \affaddr{UCLouvain Crypto Group}\\
  %     \email{olivier.pereira@uclouvain.be}
\and  % use '\and' if you need 'another row' of author names
%\alignauthor
  \email{firstname.lastname@uclouvain.be}
}
\maketitle

\section{Revisiting protocol flexibility}
\label{sec:rev_prot_flex}
Internet technologies have been designed from guidelines like the
robustness principle also known as Postel’s law~\cite{rfc-tcp}. Jon
Postel’s law is described as: “Be conservative in what you do, be
liberal in what you accept from others.”  Fundamentally, it advises
protocol designs to be tolerant with what they accept from the other
peers. In practice, this law enables protocols to be forward
compatible, meaning that they will be tolerant with a future version
of the same protocol, making it possible to avoid unrecoverable errors
when peers do not use the same protocol version. The Tor routing
protocol naturally implements the robustness principle when processing
data and control information, and it shows to be of crucial importance
within a distributed and volunteer-based network that can be composed
of nodes running many different versions of the protocol.

The robustness principle is elegant and straightforward to turn into a
practical implementation. However, its security implications are often
ignored.  Rochet and Pereira~\cite{rochet2018dropping} showed that the
robustness principle could be exploited to convey information between
malicious relays, find the guard relay used by a particular onion
service or apply other various attacks. Furthermore, real attacks were
observed in the history of the Tor network with techniques exploiting
protocol robustness~\cite{relayearly}, raising awareness over the role this
principle can have into efficiently implementing well-known theoretical attacks,
like end-to-end traffic confirmation.

  We propose to take a step back and wonder how the robustness
  principle could be revisited to support security requirements.  Our
  goal would be to define a software architecture that offers the
  benefits of the robustness principle (i.e., efficient network
  services despite the presence of various software versions), while
  also guaranteeing that this robustness cannot be exploited by making
  sure that it is only used to support authentic evolutions of the
  protocol specification.
  We start to describe two families of usage scenarios: \textbf{1) lightweight
    updates for distributed systems: flexibility without Postel's
    law}; \textbf{2) Custom Internet Privacy.}
  Then, we give an overview of the software architecture we are developing to achieve these scenarios.

\section{Flexible Anonymous Netwok} \label{sec:fan}
% Should explain how to apply the previous section to Tor

We call FAN, for Flexible Anonymous Network, an anonymous network architecture
able to transparently change its behavior for one or many users without having
to restart relays or perturbing other user connections while proceeding to add,
remove or modify protocol features. A FAN is achieved using what we call
``Protocol Plugins'', a novel technology which will be further described in
Section~\ref{sec:protocol_plugins}. We now describe two scenarios of our ongoing
research that are made possible with a FAN architecture.

\textbf{Lightweight updates.}  The first research question we consider is how to
use Protocol Plugins to build a flexible Tor network which does not blindly
comply with the robustness principle. Answering this research question would put
Postel's law behind us and may solve Tor's security issues related to it, as
well as offering new perspectives for feature deployment. A first perspective
would be to offer Tor developers a method to propagate lightweight updates in a
snap of fingers to all Tor users (relays, onion services, and Tor clients).
Lightweight updates would offer flexibility without blindly following Postel's
law, since any new feature can be added to the protocol through Protocol Plugins
and later be part of the codebase once the relay operator bumps Tor's version
through regular packaging management but, still, any deviation from the baseline
protocol specification would need to be properly authenticated. The plugins are
written in the same high-level language as Tor; they can easily be merged inside
the main codebase for people running the up-to-date Tor version, leaving
developers only to workout merge flows to manage plugins with the main codebase
during development. To manage, download, delete and verify plugins authenticity,
lightweight updates could follow TUF~\cite{samuel2010survivable}'s proposal,
(the update framework), an already widely used framework for secure updates
aimed to survive keys compromise.
%In our research, we will roll out a simple solution
%and leave this engineering problem for a proper implementation, if any.

\textbf{Custom Internet privacy.} 
An exciting direction for a FAN architecture applied to Tor would be to let Tor
users and onion services to plug code in their own Tor circuits. This could
offer the opportunity for each user to benefit from the right 
anonymity/performance trade-off during their transient usage of the Tor network. For
example, a Tor user could decide to plug a specific padding scheme over its
middle relay while visiting an onion service. On the relay, the protocol
extension would be ephemeral and only valid for this specific connection. This
research direction envisions a new type of architecture for a distributed
anonymous communication network, where setting the communication protocol to use
at a given time would be itself private, meaning that the attacker would have
the additional difficulty to apprehend which type of privacy-preserving
technique is used to deanonymize the protected data flow.

However, such remote code injection capability is also potentially dangerous for
the network performance, security, stability and for users' privacy. We designed~\cite{pquic}
a plugin management system that bears similarity to Certificate Transparency
which allows a way to distribute adequate proofs that the plug-in enforces
some claimed security properties. According to this design, relays, and end-users
would only accept plugins with valid proofs that the code is consistent with some
properties that they require to be guaranteed.

\section{Protocol Plugins} \label{sec:protocol_plugins}
% Should explain the general architecture
We suggest redesigning anonymous communication implementations such as
Tor to make them flexible through Protocol Plugins, which are pieces of code that
are executed inside a userland VM (yet with comparable performance to native
execution) in response to a particular event.

\begin{figure}[h] \centering
  \includegraphics[width=\columnwidth]{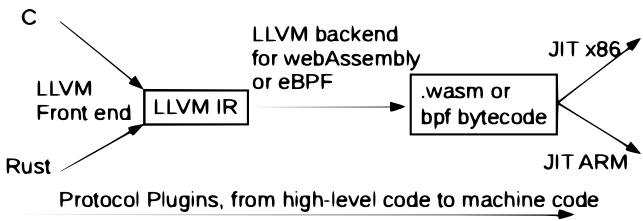} \caption{Protocol
    plugin lifecycle} \label{fig:lifecycle} \end{figure}

We design Protocol Plugins from high-level
languages such as C or Rust and we target a bytecode representation of a
conceptual machine using the LLVM compiler Clang. In our experiments, we looked
into eBPF's virtual machine and WebAssembly's virtual machine. eBPF's virtual
machine has been designed to run \bpf bytecode inside the Linux kernel, like
mini-programs which can be started in response to a particular event inside the
kernel. This \bpf bytecode is JITted to machine code and offers comparable to
native performance while executing dynamically plugged programs. On the other
hand, webAssembly has been designed to offer an alternative to JavaScript
regarding performance on critical tasks and to offer a higher diversity of
languages for the web Platform since .wasm bytecode can be obtained from the
compilation of diverse high-level languages. Figure~\ref{fig:lifecycle} offers a
representative view of the process to build Protocol Plugins. Both \bpf and
wasm's execution can be emulated during development, which makes debugging a lot
more comfortable but still translates to machine code in production.
WebAssembly in its bytecode format has nothing dependent to the Web, nor eBPF
has dependencies to the Linux kernel. Both are only instructions in a low-level
format that follows a given conceptual machine architecture.

%\COP{I significantly rewrote the paragraph below. Please check.}
Protocol Plugins are, in our design, embedded in Tor's critical path
when handling data cells and would be called for any protocol
operation that would not follow the native implementation.  We
distinguish between the operations handled by the native code which
each relay runs, depending on their Tor version, and the operations
handled by protocol plug-ins as default operations and non-default
operations (what is default or non-default then depends on the
specific Tor version installed on a node).

Non-default operations call Protocol Plugins deployed on relays in
order to follow the up-to-date version of the Tor Routing Protocol.
If no Protocol Plugin can handle the data, nor the default operation,
it means that the information received does not match the protocol's
specifications. In consequences, the relays could apply the most
efficient error-handling mechanism to protect users' privacy, such as
killing the Tor circuit and reporting the error if needed. Protocol
Plugins are a few KBs files containing the bytecode produced by the
LLVM compiler. Portable by design, Protocol Plugins can be distributed
to relays independently from their system architecture and run inside
optimized sandboxes with comparable to native performance once JITed
to machine code.

We made a prototype implementation that is already working on the
receiving side of plugins and allows plugging and executing JITed \bpf
bytecode in response to Tor's protocol events inside a sandboxed
userland VM. New protocol operations are plugged to Tor's main code in
less than 1ms on a regular laptop (8th Gen Intel, 2666Mhz DDR4 and NVMe SSD),
which accounts for the load of the plugin from the disk and the bootstrap of the
user space virtual machine used to sandbox the plugin. The current implementation of the virtual machine allows exporting
functions from the host application (Tor in this case) at compile time (e.g.,
helper functions). For more dynamical needs, like accessing some internal Tor
data from a plugin, we defined an interface between the host application and the
plugins in order to allow the plugin to get/set Tor's internal state since, by
design, we cannot dereference pointers outside of the space allocated for the
virtual machine. Data access capabilities can be given to the virtual machine
running the plugin, including Tor's internal data or even external access, like
authorizing a system call to open a particular file (e.g., writing logs) but
nothing else.  Tor's code calls the plugin in response to some particular event,
like the reception of a cell that includes an unhandled protocol feature within
its header. In that case, the cell would be transmitted to the plugin handling
that particular protocol feature.  Any other application-defined event can also
call plugins.
%\COP{As many changes (e.g., a big fix in an encryption algorithm)
  %will probably not be observable through changes in cell format, does this mean
  %that, somehow, any protocol feature that could be pluginized should start
  %having a kind of versioning mechanism that would allow detecting whether it is
  %up-to-date or whether a plugin should be used instead?  }

\section{A new paradigm?}

Anonymous communications is one of many fieds that could benefit from protocol
plugins. Protocol plugins may serve other goals, such as advancing the arm race
in censorship resistance (e.g., exploring how an authorized protocol or
application could hide another protocol from the censor through plugins),
advancing deployment's speed of new transport protocols or
extensions~\cite{pquic}, advancing the Internet's control plane
interoperability, etc. 
%\COP{What is the ``control plane''? Google seems to have
  %no idea. }

% The following two commands are all you need in the initial runs of your .tex
% file to produce the bibliography for the citations in your paper.
\bibliographystyle{abbrv} {\small
\bibliography{sig-alternate-hotpets}}  % sigproc.bib
%is the name of the Bibliography in this case You must have a proper ".bib" file
%and remember to run: latex bibtex latex latex to resolve all references
%
\balancecolumns

%\balancecolumns
\end{document}